\documentclass[aps,prl,twocolumn]{revtex4}

\usepackage{epsfig}
\usepackage{color}
\usepackage{mhchem}
\usepackage{graphicx}
\usepackage{color}
\usepackage[makeroom]{cancel} 
\usepackage[normalem]{ulem} 

\usepackage{amssymb,amsfonts,amsmath}
\usepackage{amssymb,amssymb,graphics,graphicx,amsfonts,amsmath,empheq}
\usepackage{pdfpages}

\newcommand{\bF}{\mathbf{F}}

\newcommand{\br}{\mathbf{r}}

\newcommand{\bff}{\mathbf{f}}

\newcommand{\bn}{\mathbf{n}}

\newcommand{\bbr}{\mathbf{r}}

\newcommand{\sep}{ \ \ \ , \ \ \ }

\newcommand{\beq}{\begin{equation}}
\newcommand{\eeq}{\end{equation}}
\newcommand{\beqn}{\begin{eqnarray}}
\newcommand{\eeqn}{\end{eqnarray}}
\newcommand{\pp}{\partial}
\newcommand{\dd}{{\rm d}}
\newcommand{\ee}{{\rm e}}
\newcommand{\eq}{Eq.~}

\newcommand{\fig}{Fig.\ }

\newcommand{\cO}{{\cal O}}

\newcommand{\cR}{{\cal R}}

\newcommand{\la}{\langle}

\newcommand{\ra}{\rangle}

\newcommand{\vnab}{{\bf \nabla}}

\def\di{\partial_i}
\def\dj{\partial_j}
\def\djj{\partial_{jj}}
\def\dij{\partial_{ij}}
\def\djk{\partial_{jk}}
\def\dx{\partial_x}
\def\dy{\partial_y}
\def\dxx{\partial_{xx}}
\def\dyy{\partial_{yy}}
\def\dxy{\partial_{xy}}

\def\dk{\partial_k}

\def\pk{n_k}

\begin{document}
\title{Scaling behaviour of non-equilibrium planar $N$-atic spin systems under weak fluctuations}


\author{Pablo Sartori}
\affiliation{Simons Center for Systems Biology, School of Natural Sciences, Institute for Advanced Study, Einstein Drive, Princeton, NJ 08540, U.S.A.}
\author{Chiu Fan Lee}\email{c.lee@imperial.ac.uk}
\affiliation{Department of Bioengineering, Imperial College London, South Kensington Campus, London SW7 2AZ, U.K.}
\date{\today}

\begin{abstract}
Starting from symmetry considerations, we derive the generic hydrodynamic equation of non-equilibrium $XY$ spin systems with $N$-atic symmetry under weak fluctuations. Through a systematic treatment we demonstrate that, in two dimensions, these systems exhibit two types of scaling behaviours. For $N=1$, they have long-range order and are described by the flocking phase of dry polar active fluids. For all other values of $N$, the systems exhibit  quasi long-range order, as in the equilibrium $XY$ model at low temperature.
\end{abstract}

\maketitle

Like the classification of chemical elements in the periodic table, categorising dynamical systems by their distinct scaling behaviours is of fundamental importance in physics. In such a categorisation, symmetries and conservation laws play the role of proton number in the periodic table. For  equilibrium systems, symmetries constrain the form of the Hamiltonian \cite{hohenberg1977theory}; while for non-equilibrium systems, symmetries directly constrain the form of the equations of motion, which in general can not be derived from a Hamiltonian. Therefore, imposing a set of symmetries on non-equilibrium systems  can be  less restrictive than doing so on equilibrium systems. As a result, phenomena impossible in equilibrium can occur under non-equilibrium conditions.

One example of the above is the breaking of a global continuous symmetry in two dimensions. In thermal systems such a phenomenon is forbidden by the Mermin-Wagner-Hohenberg theorem \cite{Mermin1966,hohenberg_pr67}. However, this type of symmetry breaking was shown to occur in non-equilibrium models inspired by flocking of animals \cite{vicsek1995novel}. In particular, the global rotational symmetry of a self-propelled $XY$ model can be broken in two dimensions.  The resulting ordered phase, hereafter referred to as the flocking phase, exhibits long-range order that belongs to a universality class first described by Toner and Tu \cite{toner_prl95, toner1998flocks}. But despite this celebrated example, the emergence of novel universal behaviour due to  non-equilibrium dynamics is not the norm \cite{chen_natcomm16, Chen2018}. For example, the ordered phase of an incompressible polar active fluid in two dimensions belongs to the universality class of equilibrium planar smectics \cite{chen_natcomm16}. Hence, it remains unclear how important the equilibrium constraints are with regards to the universal behaviour of a system.

\begin{figure}\label{fig:scheme}
\includegraphics{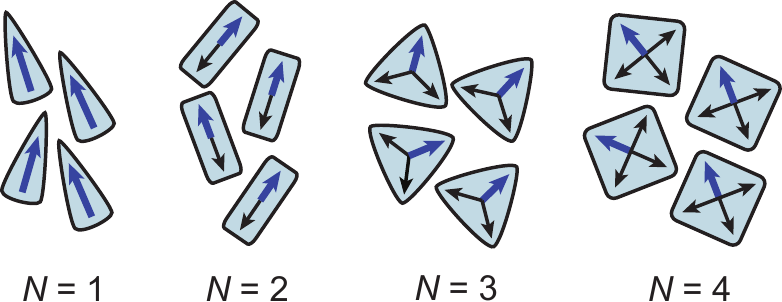}
\caption{
 {\it Schematic representation of the $N$-atic symmetry.} In this microscopic depiction the arrows represent the  $N$-atic  spins for $N=1,2,3,$ and 4. The thick blue arrow represents the local direction of the field ${\bf n}$.}
	\label{fig:pic}
\end{figure}

To shed further light on this question, we study a class of planar non-equilibrium $XY$ spin systems (see  \cite{janssen_zpb86, leung_jstatphys86, bassler_prl94, solon2013revisiting}  for  the Ising-type) on which we impose the $N$-atic symmetry. The $N$-atic symmetry is a discrete  symmetry by which a rotation of the spin's direction by an angle $2\pi/N$ leaves the system invariant  (see \fig \ref{fig:pic}). Familiar examples of $N$-atic systems are nematic liquid crystals  ($N=2$) \cite{degennes_b95}, and hexatic systems ($N=6$) in two dimensional fluids \cite{halperin_prl78, nelson_prb79} and polymerised membranes \cite{nelson_jp87, Lubensky1992}. More exotic $N$-atic symmetries, such as the tetratic symmetry ($N=4$), have also been studied theoretically \cite{Manyuhina2015}. Here, we  characterise the scaling behaviour of this class of systems in two dimensions in the limit of weak fluctuations. We focus on planar systems  because it is where the Mermin-Wagner-Hohenberg theorem is broken by the non-equilibrium dynamics of active fluids \cite{toner_prl95, toner1998flocks}.
 We will show that these systems are described by only two distinct types of hydrodynamic behaviours: for the polar case ($N=1$), the non-equilibrium system belongs to the aforementioned flocking phase \cite{vicsek1995novel, toner_prl95, toner1998flocks}; while for all other cases ($N\ge2$), the system exhibits only quasi-long range order similar to  the  equilibrium $XY$ model in two dimensions  \cite{kosterlitz1973ordering}. 
\newline

{\it  Equations of motion from symmetry.} We consider a generic dissipative system with a single hydrodynamic vector field $\bn(\bbr)$. To ease exposition, we will first  elucidate the general form of the equation of motion, and then focus on the planar case with $N$-atic symmetry. We assume that the spins tend to align so that in the weak fluctuation limit, the system is ``almost ordered''. The corresponding microscopic picture amounts to picking suitable spin directions (blue arrows in \fig \ref{fig:pic}), the vector field $\bn(\bbr)$ then corresponds to the mesoscopic average orientation of these chosen spins. We further assume that  amplitude fluctuations in $\bn$ are negligible, which amounts to the norm of $\bn$ being  a constant, which we set to one. In other words, we assume that under weak fluctuations, the system's scaling behaviour is dominated by the phase, but not amplitude, fluctuations.

Given the above, the  generic equation of motion is of the form 
\begin{align}\label{eq:n}
\pp_t \bn =  \bF(\bn, \nabla \bn, \ldots) -\lambda \bn + \bff
\end{align}
where $\bF$ is a function of the field $\bn$ and its spatial derivatives, ${\bff}$ is a noise term, and $\lambda$ is a deterministic Lagrange multiplier that imposes $|\bn|=1$. In order to preserve the norm of $\bn$,  the noise must not have a component along the local direction of the field. We thus have $f_i=P_{ij}\xi_j$, with $P_{ij}=\delta_{ij}-n_in_j$ a projection operator and $\xi_j$ the noise source (hereafter  roman subindices denote vector components, and repeated indices are summed over).
 Since we focus here on the long-time and large distance limits, we expect the coarse-grained noise term to become spatially and temporally delta-function like, and its distribution  (as long as the variance of the microscopic noise distribution does not diverge) to become Gaussian due to the central limit theorem. Therefore, we assume the noise $\xi_i$ to be Gaussian and uncorrelated, and is characterised by
\begin{subequations}
\begin{align}\label{eq:noise}
\la \xi_i (\bbr,t) \ra &=0 \quad ,\\
\la \xi_i (\bbr,t) \xi_j (\bbr',t')\ra &= 2 \Delta \delta_{ij}\delta^2 (\bbr-\bbr') \delta (t-t')\quad,
\end{align}
\end{subequations}
with $\Delta$ the noise strength, which as aforementioned we assume to be small. 	
	
The form of the generic ``force'' $\bF$ is determined by symmetry considerations alone. In this work, in addition to the $N$-atic symmetry, we assume the usual temporal, translational, rotational, and chiral symmetries in the system (see, e.g., Sect.~3.1 in \cite{lee_jpd19}). Moreover, we assume that the dimensionality of the field $\bn$ and of the embedding space $\br$ are the same. This equality allows them to couple to each other, in the sense that the dot product between the field $\bn$ and the gradient operator $\vnab$ is allowed.

We now expand $\bF$ in terms of powers of spatial derivative $\nabla$, up to and including quadratic power in $\nabla$. Since $\bF$ is a vectorial quantity and $|\bn|=1$, $\bF$ contains only terms of up to cubic order in $\bn$. Schematically, the equation of motion can be written as:
\begin{subequations}\label{eq:polar}
\begin{align}
\label{eq:B}
\pp_t \bn =& \  \sum {\bf A} \left( \nabla \nabla \bn \right)  + \sum {\bf B} \left( \bn \nabla \bn \right)\\
& + \sum {\bf C} \left(\bn \nabla \bn  \nabla \bn \right) +\sum {\bf D} \left(\bn  \bn \nabla   \nabla \bn \right)\\
& -\lambda \bn+ \bff\quad ,
\end{align}
\end{subequations}
where ${\bf A}$, ${\bf B}$, ${\bf C}$ and  ${\bf D}$ are tensors of order  four, four, six and six, respectively. In practice, the constraints of rotational symmetry and fixed norm result in  simple forms for these tensors. In the appendix we go through the straightforward calculation to demonstrate that the resulting equation of motion is
\begin{subequations}\label{eq:eom}
\begin{align}
\partial_tn_i =&\  a_1\djj n_i+ a_2\dij n_j+ b n_j\dj n_i \\
&+c_1  n_j(\dj  n_i)(\dk\pk)+ c_2  n_j(\dj \pk)(\dk n_i)\\
&+ c_3  n_j(\dj\pk)(\di\pk)+ d_1\pk n_j\dij\pk\\
&+ d_2 n_j\pk\djk n_i+ f_i\quad,
\end{align}
\end{subequations}
where $a_u$ (with $u=1,2$), $b$, $c_v$  (with $v=1,2,3$) and $d_w$  (with $w=1,2$) are model-specific coefficients that characterise the system. We emphasise that, despite the simplicity of our derivation, \eq(\ref{eq:eom}) is completely generic. In particular, it includes non-equilibrium terms such as $n_j\dj n_i$,  which is absent from the equilibrium $XY$ model since it cannot be obtained from performing a functional derivative on any functionals of $\bn$ \footnote{The term $n_in_j \pp_j n_i$ is the only term in a functional that could potentially contribute the advective term  $n_j\dj n_i$. However, $\delta (n_in_j \pp_j n_i) /\delta n_i$ will result solely in the term $n_i (\pp_j n_j)$, which can be absorbed in the Lagrange multiplier, and so the advective term remains absent.}.

So far our  treatment is valid for any dimension, and it has not incorporated the $N$-atic symmetry. Imposing the $N$-atic symmetry on planar systems requires that the dynamics remains invariant when $\bn$ is rotated around the axis perpendicular to the plane by discrete angles $\varphi_N=2\pi/N$ (for $N =1,2, \ldots$). Referring then to Eq.~(\ref{eq:n}), this symmetry amounts to the constraint
\begin{align}\label{eq:nsymm}
\cR^{\varphi_N}  \bF (\bn,\nabla\bn,\ldots) =\bF(\cR^{\varphi_N}  \bn,\nabla\cR^{\varphi_N}  \bn,\ldots)\quad,
\end{align}
where
\begin{align}
\cR^\psi \equiv \left(
\begin{array}{rr}
\cos { \psi} & \sin { \psi}\\
-\sin { \psi} & \cos { \psi}
\end{array}
\right)
\end{align}
is the two dimensional rotation matrix.  To linear order, the symmetry operation expressed by Eq.~(\ref{eq:nsymm}) is trivial. However, at the order shown  by Eq.~(\ref{eq:eom}), it results in relationships that constrain the possible values of its coefficients.  These relationships will depend on the value of $N$. We now discuss  the distinct cases.
\newline

{\it $N=1$ and the  flocking phase.} For the case $N=1$, the symmetry operation is trivial, and Eq.~(\ref{eq:eom}) remains unchanged. Since the remaining symmetries are identical to those assumed in the flocking inspired theory  of Toner and Tu, we can expect that the universal behaviour of our spin system  is the same as that of the flocking phase \cite{toner_prl95, toner1998flocks}. To show that this is indeed the case we briefly summarise  the dynamic renormalization group analysis of  Refs \cite{toner_prl95, toner1998flocks}.

First, without loss of generality, we assume that $\la \bn \ra \propto \hat{\bf x}$. Since we are interested in the limit of weak fluctuations, we expand $\bn$ as 
\begin{align}
\bn = 
\cos \theta \hat{\bf x} +\sin \theta  \hat{\bf y} \simeq (1-\theta^2/2)\hat{\bf x}+ \theta \hat{\bf y} + \cO(\theta^3)
\quad ,
\end{align}
where the scalar field $\theta(\br)$ is the angular deviation of $\bn$ from its mean value. Expressing then \eq(\ref{eq:eom}) in terms of $\theta$ to order $\cO(\theta^3)$ we obtain (see appendix):
\begin{subequations}\label{eq:theta1}
\begin{align}
\partial_t\theta =&\ \alpha_1\dyy\theta+\alpha_2\dx \theta+\alpha_3\dxx\theta+\beta\dy\theta\dx\theta\\
 &+\alpha_2\theta\dy\theta+2(\alpha_3-\alpha_1)\theta\dxy\theta+\eta
\end{align}
\end{subequations}
where, to lowest order in $\theta$, $\eta=\xi_y$,
and  $\alpha_u$ (with $u=1,2,3$) and $\beta$ are independent parameters related to the original coefficients of \eq(\ref{eq:eom}). 
 Note that $\dx \theta$ and $\theta\partial_{y}\theta$ are both derived from the non-equilibrium term $n_j\dj n_i$, and the fact that their coefficients are identical follows from rotational symmetry.

Next, we eliminate the term $\alpha_2 \pp_x \theta$ by going to the ``moving'' frame:  $x \mapsto x + \alpha_2 t$. Of the three nonlinearities in Eq.~(\ref{eq:theta1}), the $\alpha_2$ term is one order smaller in spatial derivatives than the other two,  and is hence the more relevant non-linearity. We will thus ignore the other two nonlinearities  for now,  and justify their irrelevance self-consistently {\it a posteriori}. With their omission, we arrive at the reduced equation of motion:
\begin{align}\label{eq:theta_r}
\partial_t\theta = \alpha_1\dyy\theta+\alpha_3\dxx\theta+\alpha_2\theta\dy\theta+\eta\quad .
\end{align}

We now perform the following rescalings  in \eq(\ref{eq:theta_r}): 
\begin{align}
y \mapsto \ee^\ell y \ , \ x \mapsto \ee^{\zeta\ell } x\ ,\
t \mapsto \ee^{z\ell } t \ , \ \theta \mapsto \ee^{\chi\ell } \theta\ ,
\end{align}
where $\zeta$ is the anisotropic exponent, $z$  is the dynamic exponent, and $\chi$ is commonly known as the ``roughness'' exponent \cite{barabasi_b95}. Since the only nonlinear term in (\ref{eq:theta_r}) is of the form $(\alpha_2 /2) \pp_y \theta^2$, which involves a spatial derivative with respect to $y$, the resulting Feymann diagram cannot lead to a renormalization of $\alpha_3$ nor of the noise strength $\Delta$ because neither involves a derivative with respect to $y$. In addition, $\alpha_2$ itself cannot be renormalized because \eq(\ref{eq:theta_r}) is invariant under the ``pseudo-Galilean'' transformation: $\theta \mapsto \theta +\Theta$ and  $y \mapsto y +\alpha_2 \Theta t$  for any arbitrary constant $\Theta$.  As a result, the flow equations for these coefficients are:
\begin{subequations}\label{eq:floweq}
\begin{align}
\frac{ \dd  \ln \alpha_1}{\dd \ell} &= z- 2\zeta
\\
 \frac{ \dd  \ln \alpha_2}{\dd \ell} &= z- 1+\chi
 \\
 \frac{ \dd  \ln \Delta}{\dd \ell} &= z- 2\chi -\zeta -1
 \quad .
\end{align}
\end{subequations}
Note that $\alpha_1$ does get renormalized by loop diagrams. However, determining the fixed point of the above three equations already  provide enough constraints to fix the three scaling exponents:
\begin{align}
\zeta = \frac{3}{5} \sep z =\frac{6}{5} \sep \chi = -\frac{1}{5} 
\quad .
\end{align}
These are the exact  exponents that describe the scaling behaviour of the flocking phase in two dimensions \cite{toner_prl95, toner1998flocks}. Since   $\chi<0$, $\la \theta(0,t) \theta(\br,t) \ra \sim r^{2\chi}$ goes to zero at large $r \equiv|{\bf r}|$. Therefore, a true ordered phase exists for our system because in the large $r$ limit, $\la \bn(0,t) \cdot \bn (\br,t)\ra \approx {{\rm constant}}+\la \theta (0,t) \theta(\br, t) \ra$ under weak fluctuations, which remains greater than zero as $r$ goes to infinity. Using these exponents we can now verify that the other two nonlinearities in \eq(\ref{eq:theta1}) are indeed irrelevant.

We note that the flocking phase  characterised by the above exponents is known to describe the scaling behaviour of  Malthusian flocks in two dimensions \cite{toner2012birth}, and of incompressible active fluids in dimensions higher than two \cite{chen_njp18}. We have now shown that it also describes the behaviour of planar spin systems out of equilibrium. However, whether the flocking phase further describes the behaviour of compressible flocks, which it was originally devised to do \cite{toner_prl95}, remains unsettled \cite{toner_pre12}.
\newline


{\it $N=2$ and  quasi-long range order.} When $N=2$ Eq.~(\ref{eq:nsymm}) reduces to $\bF (\bn,\nabla\bn,\ldots) =-\bF(- \bn,-\nabla\bn,\ldots)$, which indicates that $\bn$ behaves as a nematic \cite{degennes_b95}. In this case only odd powers of the field can be present in \eq(\ref{eq:n}), and so $b=0$ in (\ref{eq:eom}), which implies $\alpha_2=0$ in (\ref{eq:theta1})  (see appendix). Compared to the polar case, $N=1$, the relevant nonlinearity is now absent and the universal behaviour of the system is therefore dictated by the other two nonlinearities in (\ref{eq:theta1}): $\pp_y \theta \pp_x \theta$ and $\theta \pp_{xy} \theta$.

Interestingly, an analysis of \eq(\ref{eq:theta1}) for the case $\alpha_2=0$ has already been carried in \cite{mishra2010dynamic} in the context of active nematics adsorbed on a substrate. A one-loop dynamic renormalization group analysis indicated that the two nonlinearities in (\ref{eq:theta1}) are in fact marginally irrelevant. Therefore, for $N=2$   the hydrodynamic behaviour of our system is characterized by the linear theory that is analogous to the linear theory of the equilibrium $XY$ model, and thus admits only quasi-long-ranged order. We note that in the system studied in  \cite{mishra2010dynamic},  the dynamics of the director field $\bn$, besides the imposed active motility, is based on a Landau-de Gennes free energy. In contrast, 
 the system considered here is only constrained by symmetries, and as such, our derivation of \eq(\ref{eq:theta1}) is arguably more general.
\newline

{\it Generalisation for $N>2$.} The long-range behaviour of all the remaining cases ($N>2$) can be analysed simultaneously in a straightforward manner.  First, we note that all linear terms in $\bn$ will be present, as they all trivially satisfy the symmetry condition in Eq.~(\ref{eq:nsymm}). Next, we study how the symmetry condition  affects the term $bn_j \pp_j n_i$ in Eq.~(\ref{eq:eom}). To do so, we note that the corresponding right-hand side term of Eq.~(\ref{eq:nsymm}) is transformed as
\begin{align}\label{eq:sym1}
n_j \pp_j n_i \mapsto \cR^{\varphi_N}_{jk} n_k \pp_j \left(\cR^{\varphi_N}_{il} n_l\right)
=\cR^{\varphi_N}_{il} \cR^{\varphi_N}_{jk} n_k \left( \pp_j  n_l\right)\ ,
\end{align}
while the corresponding left-hand side term is transformed as
\begin{align}\label{eq:sym2}
n_j \pp_j n_i \mapsto\cR^{\varphi_N}_{ik}  n_j \left( \pp_j  n_k\right) = \cR^{\varphi_N}_{ik}\delta_{jl}  n_j \left( \pp_l  n_k\right)
\ .
\end{align}
Because there are no other terms with the same order of derivative in Eq.~(\ref{eq:eom}), to satisfy the $N$-atic symmetry these two terms have to match. Therefore, Eq.~(\ref{eq:sym1}) and Eq.~(\ref{eq:sym2}) have to be equal for all values of $\bn$, which requires that the two order four tensors $\cR^{\varphi_N}_{il} \cR^{\varphi_N}_{jk}$ and $\cR^{\varphi_N}_{il}\delta_{kj}$ are identical. In other words, $\cR^{\varphi_N}_{ij} =\delta_{ij}$, and the term $n_j \pp_j n_i$ can only be present if $N=1$, and must be absent otherwise.

Finally, we note that the above  argument can not be applied to the $c_v$ and  $d_w$ terms separately, as the symmetry condition will generally result in relations among these coefficients that depend on the value of $\varphi_N$ \cite{manyuhina2015forming}. However, as we have seen in the $N=2$ case, even if these terms are present, they will be marginally irrelevant.  As a result,  for $N>2$, the spin systems again admit only quasi-long-ranged order.
\newline

{\it Conclusion \& Outlook.}
We have considered a generic non-equilibrium  $XY$ spin system in two dimensions with $N$-atic symmetry and characterised its scaling behaviour under weak fluctuations for all $N$. Specifically, for the $N=1$ case, the system belongs to the flocking phase, first described in the context of dry polar active fluids \cite{toner_prl95,toner1998flocks}; and for all subsequent values of $N$, the system exhibits quasi-long-ranged order similar to the equilibrium $XY$ model at low temperature.  In other words, as far as universality is concerned, the non-equilibrium nature of the system is only manifested for $N=1$. 
 We however stress that this conclusion applies only in the weak fluctuation limit of the systems considered, and there are in fact  diverse emergent behaviour in active systems (e.g., in active nematic systems) that exhibit phenomena not seen in equilibrium systems.

Our work is closely related to non-equilibrium theories of fluids.  The key parallel is that in both classes of systems the spatial dimension and the field dimension coincide. This allows  for the contraction of the indices of spatial derivatives with those of the field, giving rise to a richer mathematical structure in Eq.~(\ref{eq:eom}). Looking ahead, it would  be interesting to consider the scaling behaviour of other  non-equilibrium spin systems, such as the Potts model, where some of the components are coupled to the spatial dimensions. Another promising direction would be to study how the non-equilibrium dynamics would affect the shape of the systems if the spins are coupled to the metric tensor of the underlying space \cite{Lubensky1992, Manyuhina2015}.  More generally, our work shows that for a large class of {spin} systems ($N\ge2$), breaking the equilibrium constraints does not result in novel  hydrodynamic behaviour. Therefore, a fundamental open question is:   What are the underlying conditions for the emergence of novel phases out of equilibrium, as  in the polar ($N=1$) case?

\appendix
    \section{Appendix: Derivation of the general equation of motion.} 
	As discussed in the main text, the deterministic component of the equation of motion of $\bn$, with the constraint $|\bn|=1$, and up to terms of order $\cO( \vnab^2)$, is:
\begin{subequations}\begin{align}
		\pp_t \bn =& \  \sum {\bf A} \left( \nabla \nabla \bn \right)  + \sum {\bf B} \left( \bn \nabla \bn \right)\\
&+ \sum {\bf C} \left(\bn \nabla \bn  \nabla \bn \right) +\sum {\bf D} \left(\bn  \bn \nabla   \nabla \bn \right)\\
&-\lambda \bn \quad ,
\end{align}\end{subequations}
where ${\bf A}$, ${\bf B}$, ${\bf C}$ and  ${\bf D}$ are tensors of order four, four, six and six, respectively; and $\lambda$ is a Lagrange multiplier that enforces the constraint on the norm of $\bn$.

We now write down all possible terms corresponding to each of these sums,  with the corresponding small letters $a_u$, $b_v$, $c_w$, and $d_z$. We use index notation, so that $\bn$ is denoted by $n_i$, $\nabla$ is denoted by $\partial_i$, and repeated indices are summed over:
\begin{subequations}
\begin{align}
{ {\bf A}:} 
\ \{&\dj\dj n_i, \quad\di \dj n_j\}\\
{ {\bf B}:} 
\ \{&n_i\dj n_j,\cancelto{0}{\quad n_j\di n_j},\quad n_j\dj n_i\}\\
{ {\bf C}:}
\ \{&n_i(\dj  n_j)(\dk \pk), \quad n_i(\dj\pk)(\dj\pk),\\
&n_i(\dj\pk)(\dk n_j),\quad n_j(\dj  n_i)(\dk\pk),\\
&n_j(\dj \pk)(\dk n_i),\quad  n_j(\dj\pk)(\di\pk),\\
&\cancelto{0}{ n_j(\di n_j)(\dk\pk)},\quad \cancelto{0}{ n_j(\di\pk)(\dk n_j)},\\
&\cancelto{c_6}{ n_j(\di\pk)(\dj\pk)},\quad\cancelto{0}{ n_j(\dk n_j)(\dk n_i)},\\
&\cancelto{0}{ n_j(\dk n_j)(\di\pk)}, \cancelto{0}{ n_j(\dk n_i)(\dk n_j)},\\
&\cancelto{c_5}{ n_j(\dk n_i)(\dj\pk)}\}\\
{ {\bf D}:}
\ \{&\cancelto{a_1}{n_kn_k\dj\dj n_i},\quad n_in_k\dj\dj n_k,\\
&\cancelto{a_2}{n_kn_k\di\dj n_j}, \quad {n_kn_j\di\dj n_k},\\
& n_i\pk\dj\dk n_j,\quad n_j\pk\dj\dk n_i\}
\end{align} 
\end{subequations}
where the null terms cancel because of the constraint $\bn^2=1$. This constraint also makes several terms equal. After taking all the cancellations and equalities into account, we have
\begin{subequations}
	\label{eq:app}
\begin{align}
\partial_tn_i =& -\lambda n_i+a_1\djj n_i+a_2 \dij  n_j+b_1n_i\dj n_j\\
&+b_2 n_j\dj n_i+c_1n_i(\dj  n_j)(\dk \pk)\\
&+c_2 n_i(\dj\pk)(\dj\pk)+c_3  n_i(\dj\pk)(\dk n_j)\\
&+c_4  n_j(\dj  n_i)(\dk\pk)
+c_5  n_j(\dj \pk)(\dk n_i)\\
&+c_6  n_j(\dj\pk)(\di\pk)
+d_1 n_in_k\djj n_k\\
&+d_2\pk n_j\dij\pk+d_3 n_i\pk\djk n_j\\
&+d_4 n_j\pk\djk n_i
\quad .
\end{align}
\end{subequations}
This equation can be simplified further, by noting that all terms $\propto n_i$ can be directly absorbed in the Lagrange multiplier that enforces the norm constraint.  Finally, the norm-preserving equation of motion of $\bn$ becomes
\begin{subequations}
\label{eq:eom_n_all}
\begin{align}\label{eq:eom_n_lam}
\partial_tn_i =& \ a_1\djj n_i+a_2 \dij n_j+b_2 n_j\dj n_i\\
&+c_4  n_j(\dj  n_i)(\dk\pk)+c_5  n_j(\dj \pk)(\dk n_i)\\
&+c_6  n_j(\dj\pk)(\di\pk)+d_2\pk n_j\dij\pk\\
&+d_4 n_j\pk\djk n_i
\ ,
\end{align}
\end{subequations}
 which, after relabelling the indices of the coefficients, is exactly the deterministic component of Eq.~(\ref{eq:eom}) in the main text.


To arrive at \eq (\ref{eq:theta1}) in the main text, we expand $\bn$ as $\cos \theta \hat{\bf x} + \sin \theta \hat{\bf y} \approx (1-\theta^2/2)\hat{\bf x}+\theta \hat{\bf y}$ and focus on the $y$-component to get
\begin{subequations}
\begin{align}\label{eq:eom_ang_si}
\partial_t\theta=& \ \alpha_1\dyy\theta+\alpha_2\dx \theta+\alpha_3\dxx\theta+\beta_1\dy\theta\dx\theta\\
&+\beta_2\theta\dy\theta+\beta_3\theta\dxy\theta
\end{align}
\end{subequations}
where the coefficients relate to the those in \eq (\ref{eq:eom_n_all}) by $\alpha_1=a_1+a_2$, $\alpha_2=b_2$, $\alpha_3=a_1+d_4$, $\beta_1=c_4+c_5+c_6-a_2-d_2$, $\beta_2=b_2$, $\beta_3=2(d_4-a_2)$. Note in particular that the following relations are satisfied:  $\alpha_2=\beta_2$ and  $\beta_3=2(\alpha_3-\alpha_1)$.
    
\vspace{.1in}
\begin{acknowledgments}
{\it Acknowledgements.} We thank Jordan Horowitz and Rui Ma for useful comments on the manuscript.  P.S. is funded by the Eric and Wendy Schmidt Membership in Biology at the Institute for Advanced Study.
\end{acknowledgments}


\end{document}